\documentclass[a4paper,11pt]{article} \usepackage{pos} \usepackage{subcaption}

\title{Status of reproducibility and open science in hep-lat in 2021}

\author*[a]{Ed Bennett}

\affiliation[a]{Swansea Academy of Advanced Computing, Swansea University, Bay
  Campus, Fabian Way, Swansea SA1 8EN}

\emailAdd{e.j.bennett@swansea.ac.uk}

\abstract{As a fully computational discipline, Lattice Field Theory has the
  potential to give results that anyone with sufficient computational resources
  can reproduce, going from input parameters to published numbers and plots
  correct to the last byte. After briefly motivating and outlining some of the
  key steps in making lattice computations reproducible, this contribution
  presents the results of a survey of all 1,229 submissions to the hep-lat arXiv
  in 2021 of how explicitly reproducible each is. Areas where LFT has
  historically been well ahead of the curve are highlighted, as are areas where
  there are opportunities to do more.}

\FullConference{%
  The 39th International Symposium on Lattice Field Theory,\\
  8th-13th August, 2022,\\
  Rheinische Friedrich-Wilhelms-Universität Bonn, Bonn, Germany }

\begin{document}
\maketitle

\section{Introduction}

There has been increasing pressure from many sides in recent years to embrace
the principle of open science, and to ensure that data analyses are
reproducible. As the software used to analyse data has become increasingly
complex, it has become less and less feasible to unambiguously explain in a
traditional journal publication exactly what steps were carried out, to the
point that another researcher could reproduce them. As enabling this is one of
the key aims of academic publication, this trend poses a clear problem, which
research funders, journals, and others are looking to address.

As a purely computational field, Lattice Field Theory (LFT) stands to be more
severely affected than most fields. However, LFT is also one of the very early
adopters of many principles of Open Science. The latter fact, specifically the
ubiquitous use of freely-available preprints for written publications, has
enabled the work presented here: an analysis of all 1,229 submissions to the
arXiv~\cite{arxiv} preprint server in the hep-lat category in 2021, to assess
where the field is strong and where there remain opportunities for improvement.

\section{Terminology}

In this work, \emph{reproducible} is used as defined by the Turing Way
project~\cite{turingway}: that given the same data, another researcher can
repeat the same analysis and obtain the same results. This definition may seem
trivial at first glance, as it seems obvious that the same analysis on the same
data should give the same result; however, it is a prerequisite of the more
challenging concepts of \emph{replicability} (being able to apply the same
analysis to fresh data and obtain the same results), and \emph{robustness}
(being able to apply different analyses to the same data and obtain compatible
conclusions).

Since the mapping from human languages to programming ones is not one-to-one,
for any non-trivial software then access to the code is an essential part of
reproducibility. Similarly, for data beyond those that can be presented in a
handful of tables, access to the raw data is also needed. Removing the
possibility for human error is another way to improve reproducibility; every
manual step in a process is a place that could be done inconsistently (either
within a work, or between the original work and attempts to reproduce it). The
primary purpose of computers is to automatically perform repeated tasks in a
consistent way, so encoding manual workflows in software significantly reduces
the possibility for human error, and where errors do creep into code, then they
can be inspected and discovered after the fact.

\emph{Open science} is the movement that all research---including not only
publications, but also physical samples, data, software, and other
outputs---should be accessible to all by default (barring specific ethical or
legal constraints, for example when working with private personal data).
Theoretical particle physics pioneered the use of open-access preprints for
publications, and remains a leading field---in many disciplines preprints are
still seen by many as a novelty. (For example, in a survey of 3759
researchers~\cite{preprintsurvey}, 46\% of respondents in medical and health
sciences had never viewed or downloaded a preprint, and 63.6\% of the same
cohort had never authored one.)

Data and software published openly can (and should) be FAIR\@: specifically,
\emph{findable}, \emph{accessible}, \emph{interoperable}, and \emph{reusable}.
This is a separate consideration to the data being made available publicly---data
that will never leave a research group still benefits from new researchers to
the group being able to easily find it, and conversely it is entirely possible
to share data in un-FAIR ways. Making data and software FAIR increases the uses
that others can make of it, and hence makes it more valuable.

\section{Survey results}

\begin{figure}
  \centering
  \includegraphics[width=0.6\columnwidth]{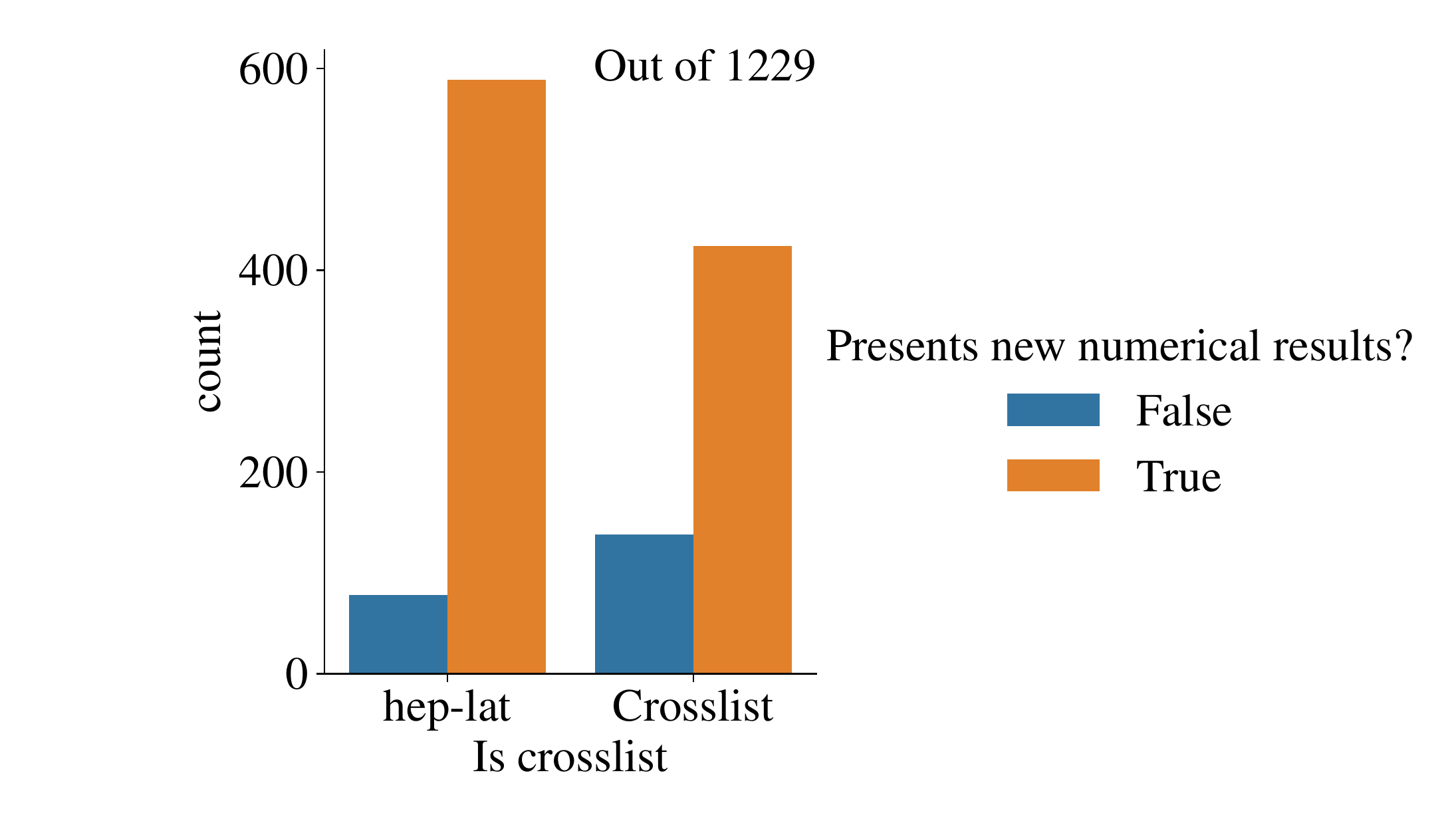}
  \caption{\label{fig:allpapers}Breakdown of submissions to the hep-lat arXiv in
    2021 indicating whether they had hep-lat as their primary arXiv or were
    crosslists, and whether they presented new numerical results.}
\end{figure}

The 1,229 submissions to the hep-lat arXiv (including those cross-listed into
the category) were surveyed for some aspects affecting the reproducibility of
the work done. The questions were answered based on a combination of full-text
search for relevant keywords and a brief skim read of the contents. The complete
survey data, and the analysis scripts used to prepare the plots presented here,
are released separately~\cite{datapackage}. The analysis was performed using
Python~\cite{python}, pandas~\cite{pandas}, Matplotlib~\cite{matplotlib}, and
seaborn~\cite{seaborn}, via a Jupyter Notebook~\cite{notebook}.

Following the principle of findability, and that a paper should include
sufficient information to be able to reproduce it without hunting elsewhere,
information not included in the preprint was not considered when completing the
survey. This excluded information added in published manuscripts but not updated
on the arXiv, and also excluded information that could be found by digging
through a collaboration's website.

Figure~\ref{fig:allpapers} shows an initial breakdown of the submissions; a
small majority had hep-lat as their primary category, with the remainder being
cross-lists. A substantial majority of both cases---1013 in all---included numerical
results (e.g.\ a plot that was not merely schematic, or numbers with
uncertainties not quoted from another publication); it is this subset that will
be the focus of the remainder of the analysis.

\begin{figure}
  \includegraphics[width=0.51\columnwidth]{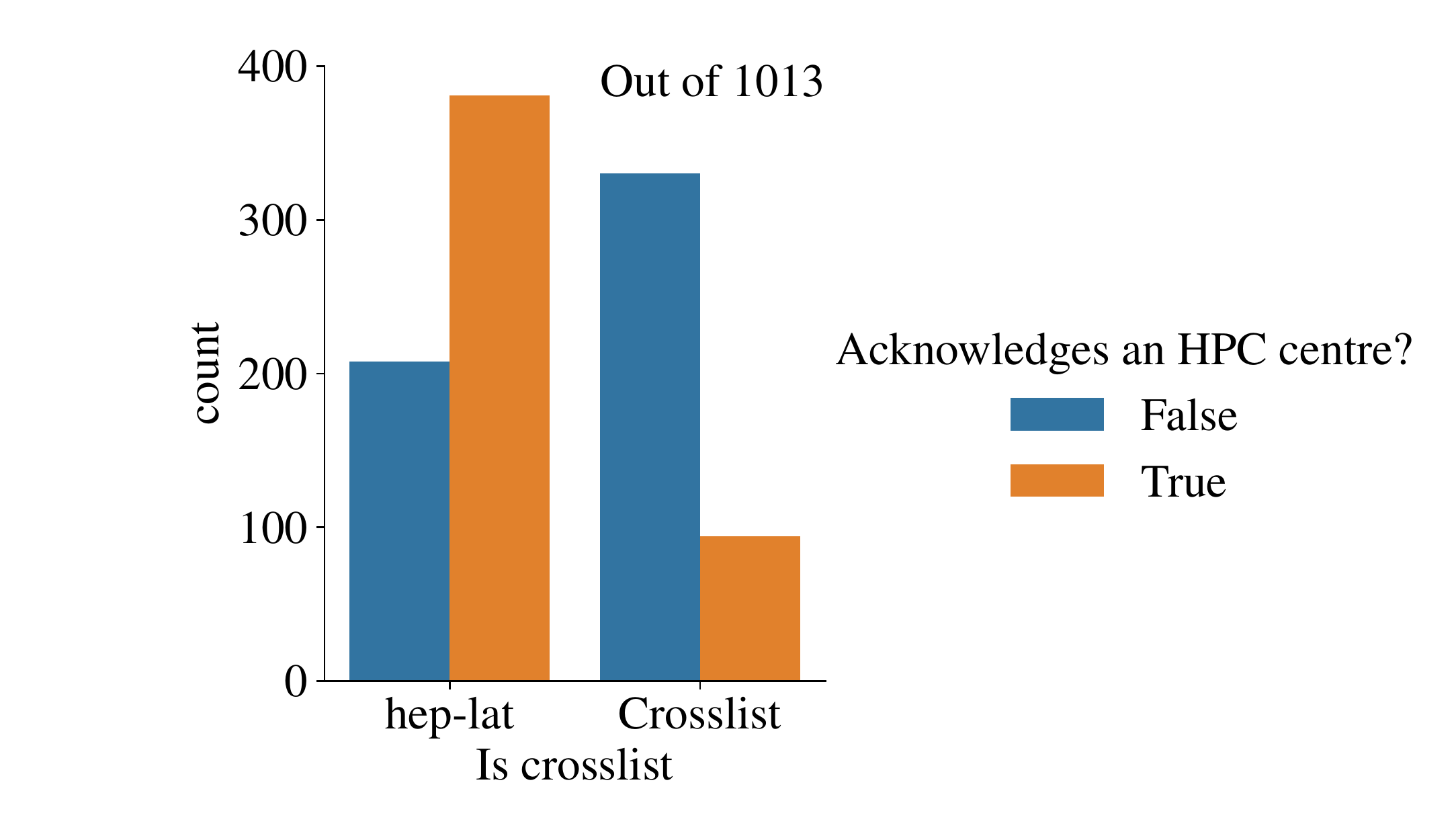}
  \hfill
  \includegraphics[width=0.47\columnwidth]{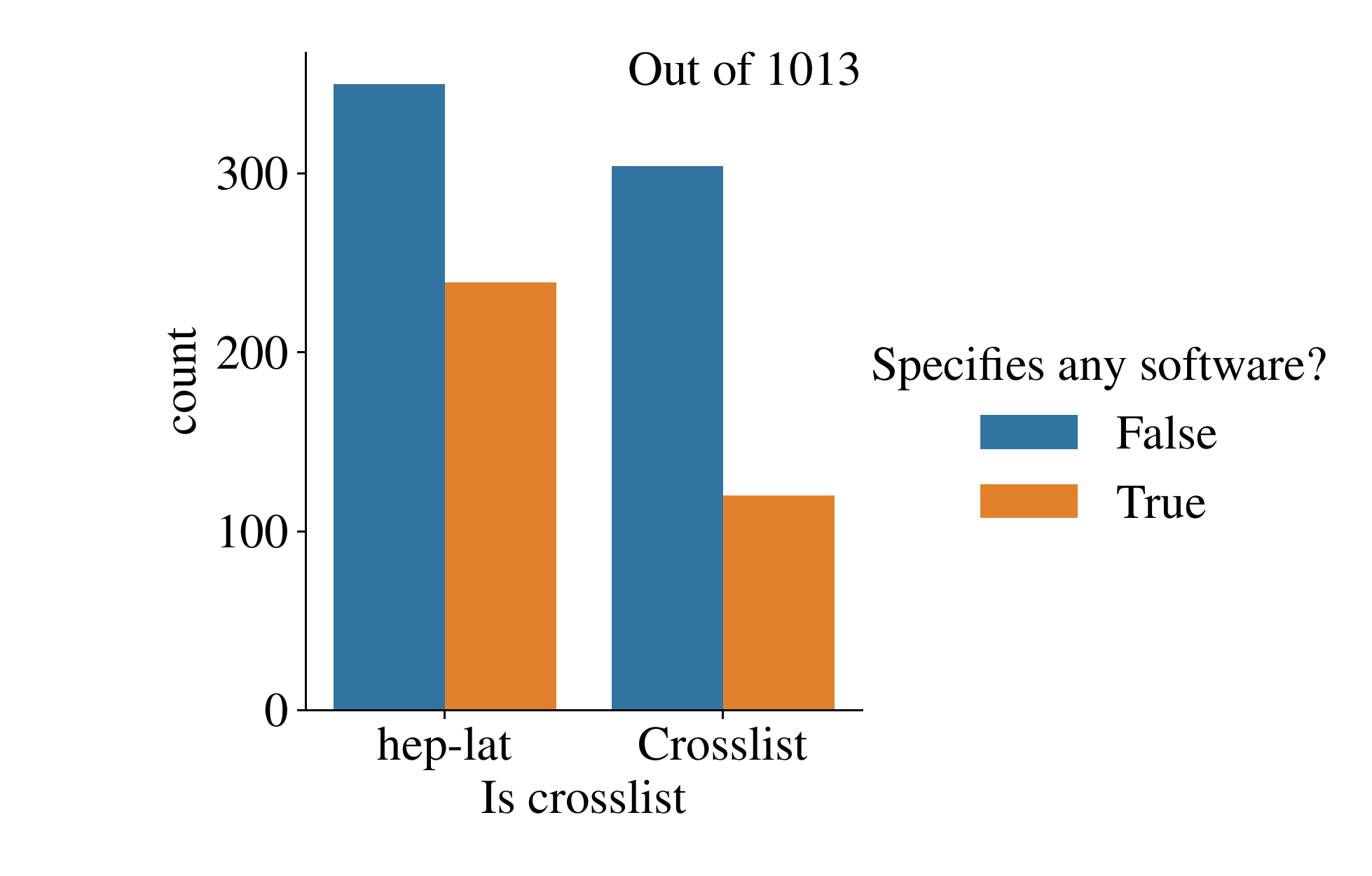}
  \caption{\label{fig:hpc-software-acknowledged}Breakdown of whether submissions
    to the hep-lat arXiv in 2021 that included new numerical results
    acknowledged an HPC resource (left panel) or specified any of the software
    used to generate their results (right panel).}
\end{figure}

Since one step towards reproducibility is to specify the software used, an
initial question to ask is what proportion of submissions do this. As a
baseline, this is compared with the proportion of submissions that acknowledge
the use of a high-performance computing (HPC) facility. These are of similar
difficulty---adding one or two sentences---and fulfil a similar purpose, in
demonstrating the utility of the software/facility and ensuring that its
development or activities continue to be funded.
Figure~\ref{fig:hpc-software-acknowledged} presents the breakdown of submissions
for these questions. The vast majority of submissions presenting new numerical
results with hep-lat as the primary arXiv (referred to hereafter as
\emph{hep-lat numerical submissions}) acknowledge the use of an HPC facility.
However, fewer than half of hep-lat numerical submissions specify or acknowledge
any of the software used. This is potentially the simplest change for authors to
make; while citing software isn't sufficient to ensure reproducibility if custom
code has been written, it significantly reduces the barrier while also
recognising software developers' contributions to the research.

The survey takes a relatively reductive view towards categorising the work done
in LFT\@: it is assumed that a typical piece of work may generate field
configurations using a Monte Carlo-like algorithm, may compute observables on
such configurations (generated as part of the same work or otherwise), and will
do some kind of statistical analysis on data resulting from one or both of the
previous two steps. (Alternative approaches including tensor networks and
quantum simulation have their software effort considered in the ``analysis''
category; their count is sufficiently small that this is not considered to
introduce significant bias.)

\begin{figure}
  \hfill
  \includegraphics[width=0.35\columnwidth]{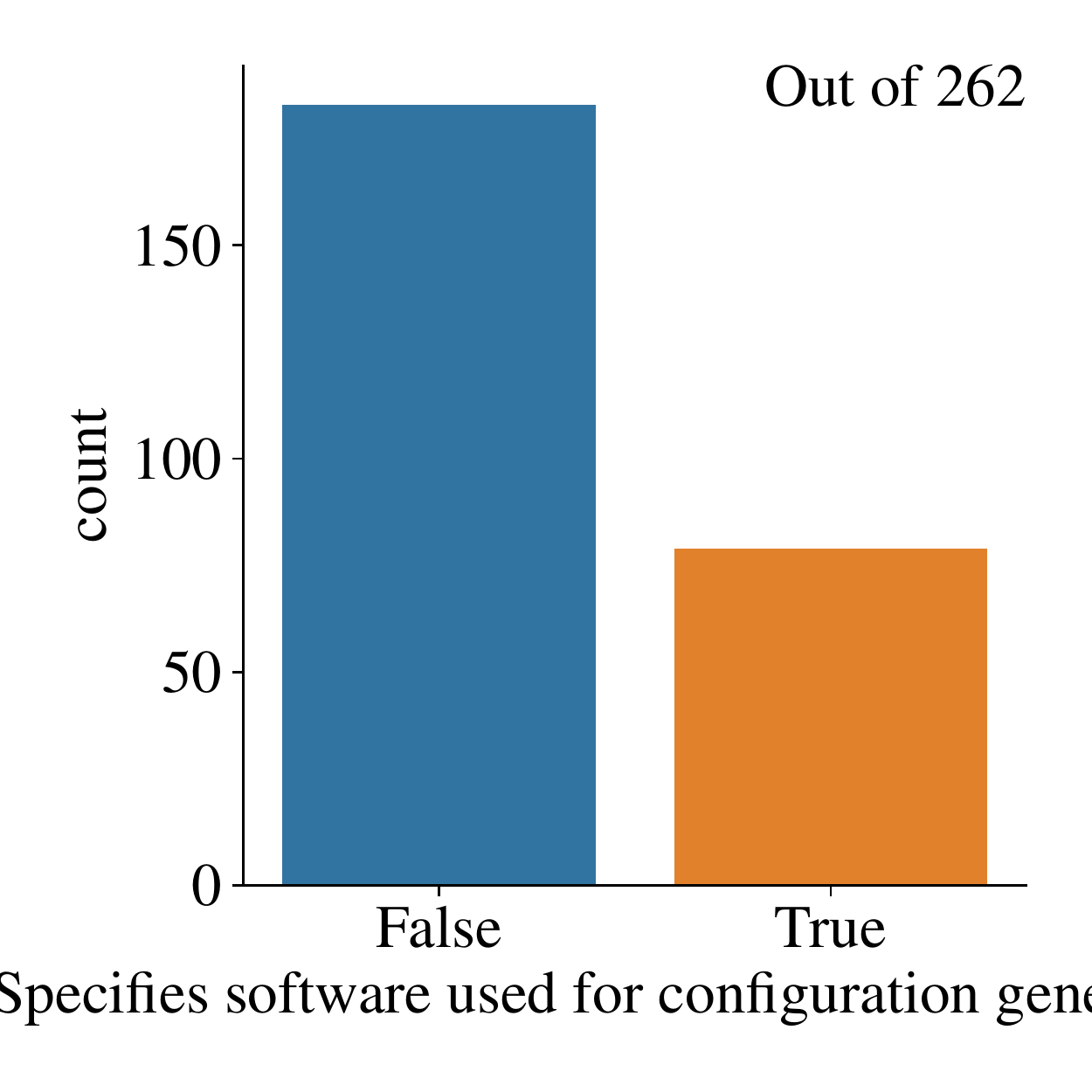}
  \hfill
  \includegraphics[width=0.35\columnwidth]{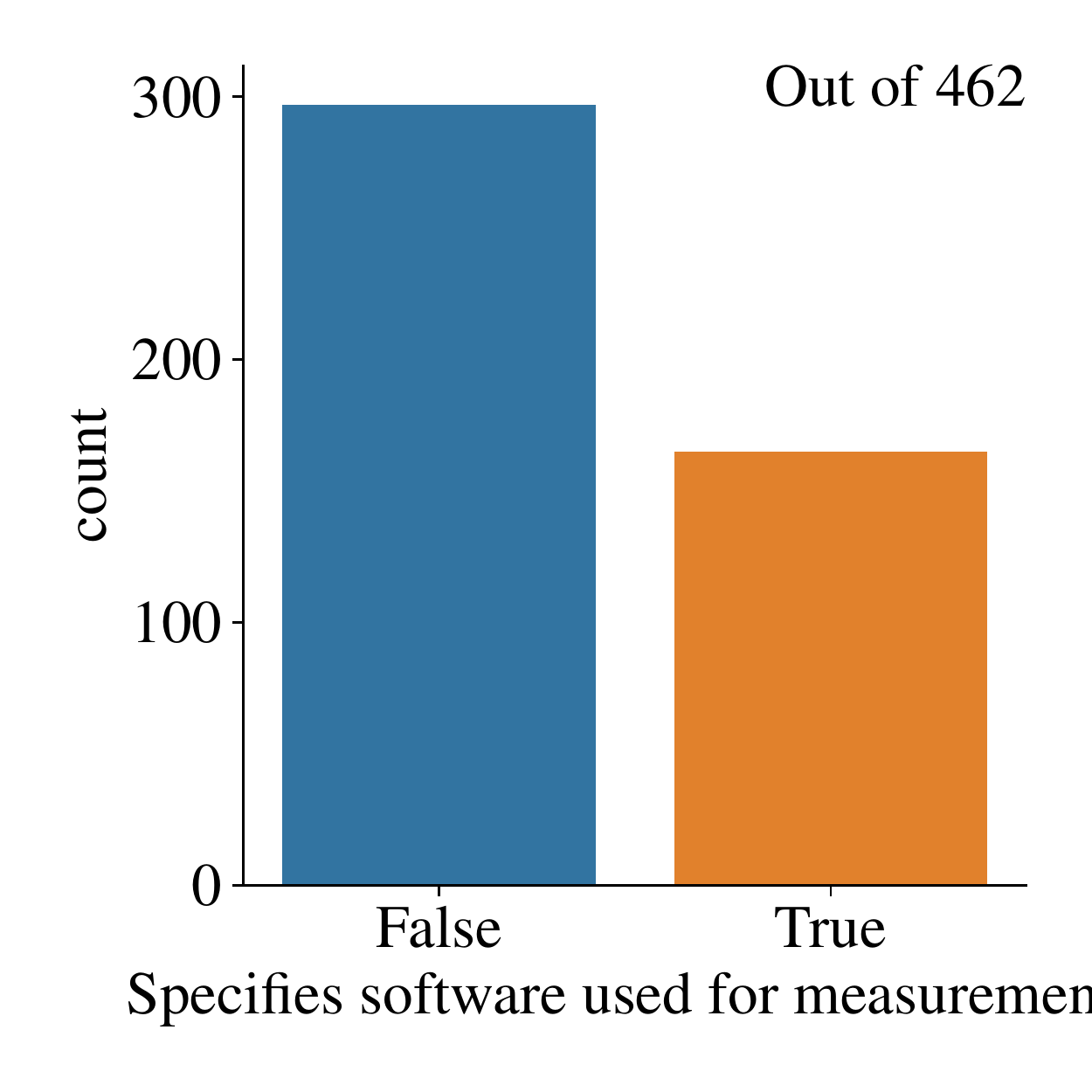}
  \hfill\hphantom{0px}
  \caption{\label{fig:specifies-config-software}Breakdown of whether submissions
    to the hep-lat arXiv in 2021 that generate new field configurations (left
    panel) or measure observables on field configurations (right panel) specify
    the software that was used to do this.}
\end{figure}

Generation of configurations is the most computationally expensive activity in
LFT, and the associated software suites have had corresponding amounts of effort
put into ensuring they are robust and efficient. Steps taken to ensure the
reproducibiliy of generation include having deterministic seeded random number
generators, and having detailed log files and storing metadata in configurations
recording what version of a piece of software was used and on what machine.
Computation of observables from these ensembles is typically the next most
expensive activity, and frequently uses some of the same software
infrastructure.

Figure~\ref{fig:specifies-config-software} shows the proportion of submissions
that specify the software used for each of these activities. Less than one in
three of the 262 submissions generating new configurations specified the
software, and a slightly larger proportion of the 462 submissions measuring
observables on configurations did.\footnote{In each case, some of the software
specified was toolkits such as Grid~\cite{grid} and Chroma~\cite{chroma}, where
significant amounts of additional code (e.g. XML or C++) is needed to perform
the work being reported; this additional code was not typically found to be
shared or specified.} Given that due to differences in normalisations and other
undocumented conventions, running two different codes with the same parameters
can give differing results, this is a barrier to reproducibility of this effort.

\begin{figure}
  \centering
\begin{minipage}{.48\textwidth}
  \centering
  \includegraphics[width=.7\linewidth]{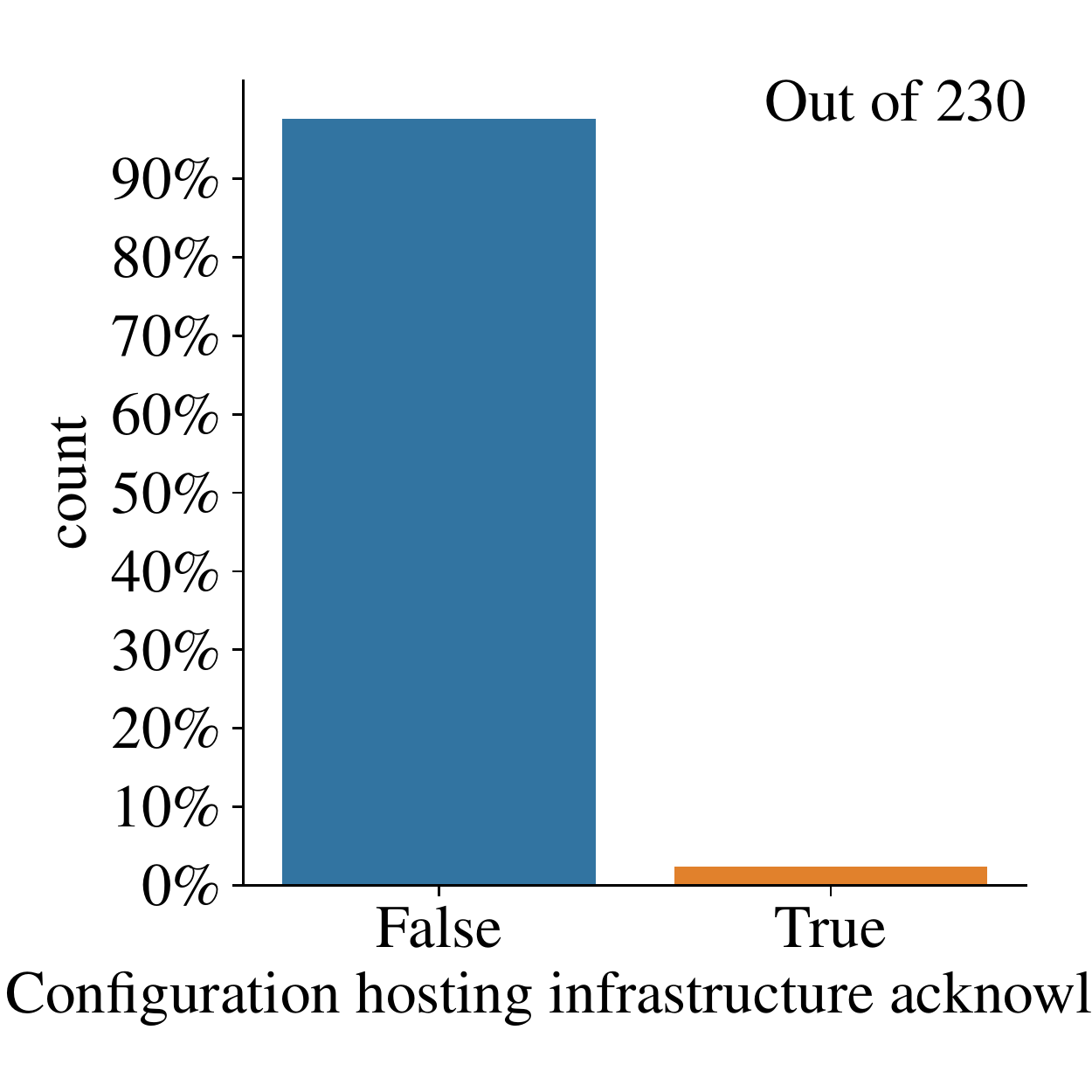}
  \captionof{figure}{\label{fig:acknowledges-cfg-hosting}Plot showing the
    proportion of submissions to the hep-lat arXiv in 2021 making use of
    existing gauge configurations that acknowledge the infrastructure used to
    host and serve them.}
  \includegraphics[width=.8\linewidth]{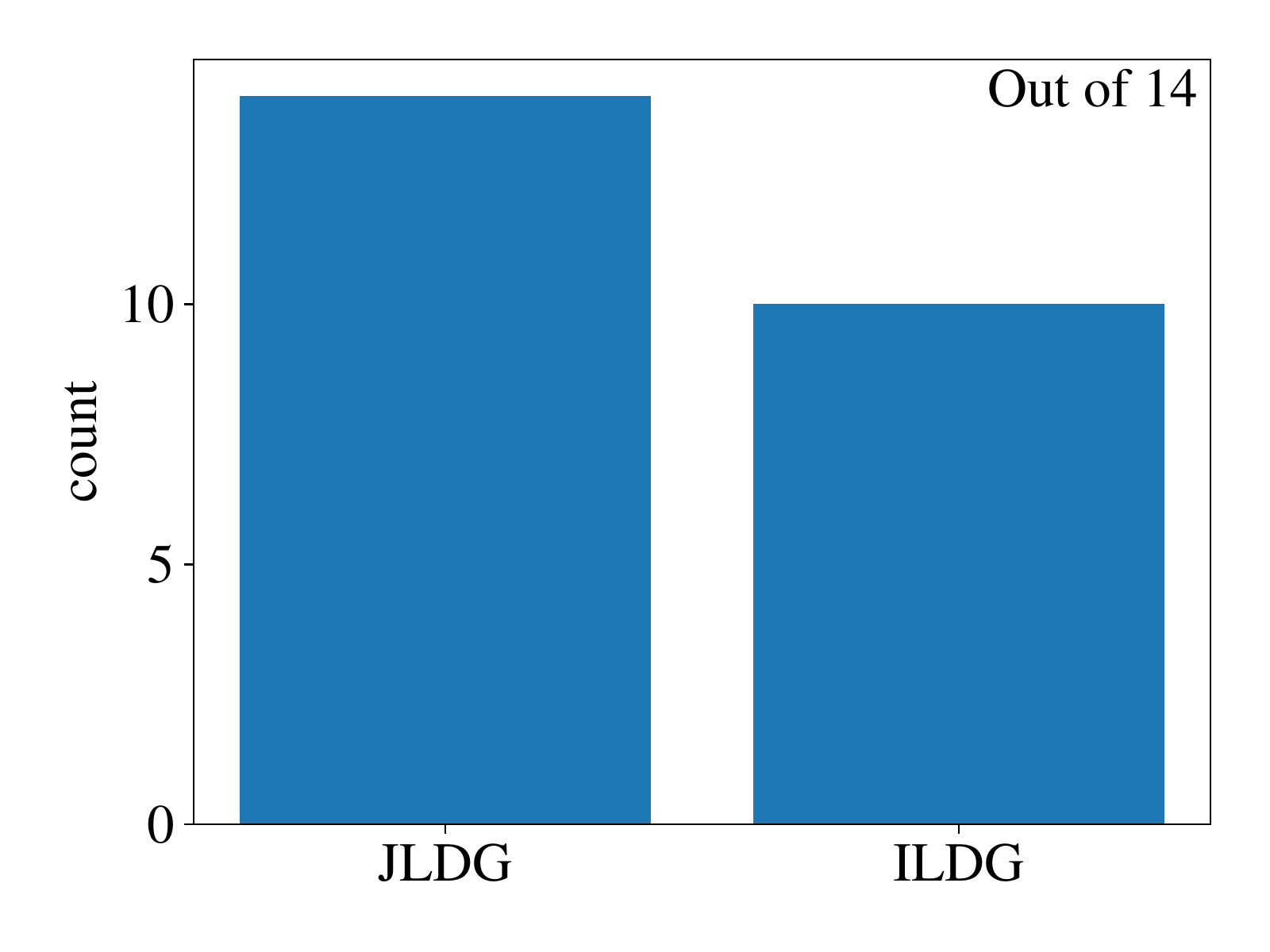}
  \captionof{figure}{\label{fig:cfg-hosting-breakdown}Breakdown of the
    infrastructure acknowledged by those submissions in
    Fig.~\ref{fig:acknowledges-cfg-hosting} that do acknowledge hosting
    infrastructure.}
\end{minipage}%
\hfill
\begin{minipage}{.48\textwidth}
  \centering
  \vspace{12pt}

  \includegraphics[width=\linewidth]{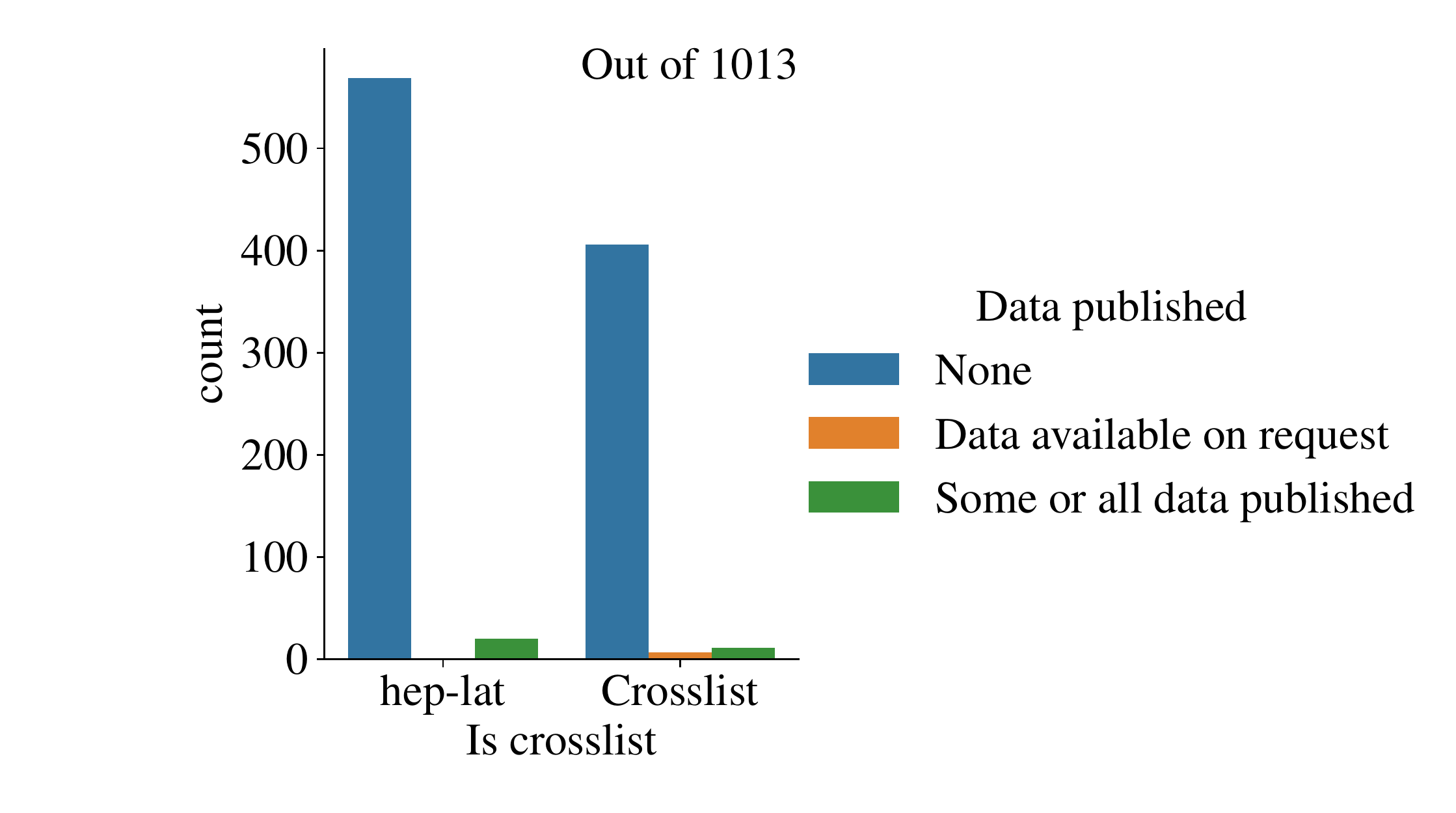}
  \captionof{figure}{\label{fig:publish-data}Breakdown of submissions to the
    hep-lat arXiv in 2021 that publish any of the data they generate.}
  \vspace{12pt}

  \includegraphics[width=.9\linewidth]{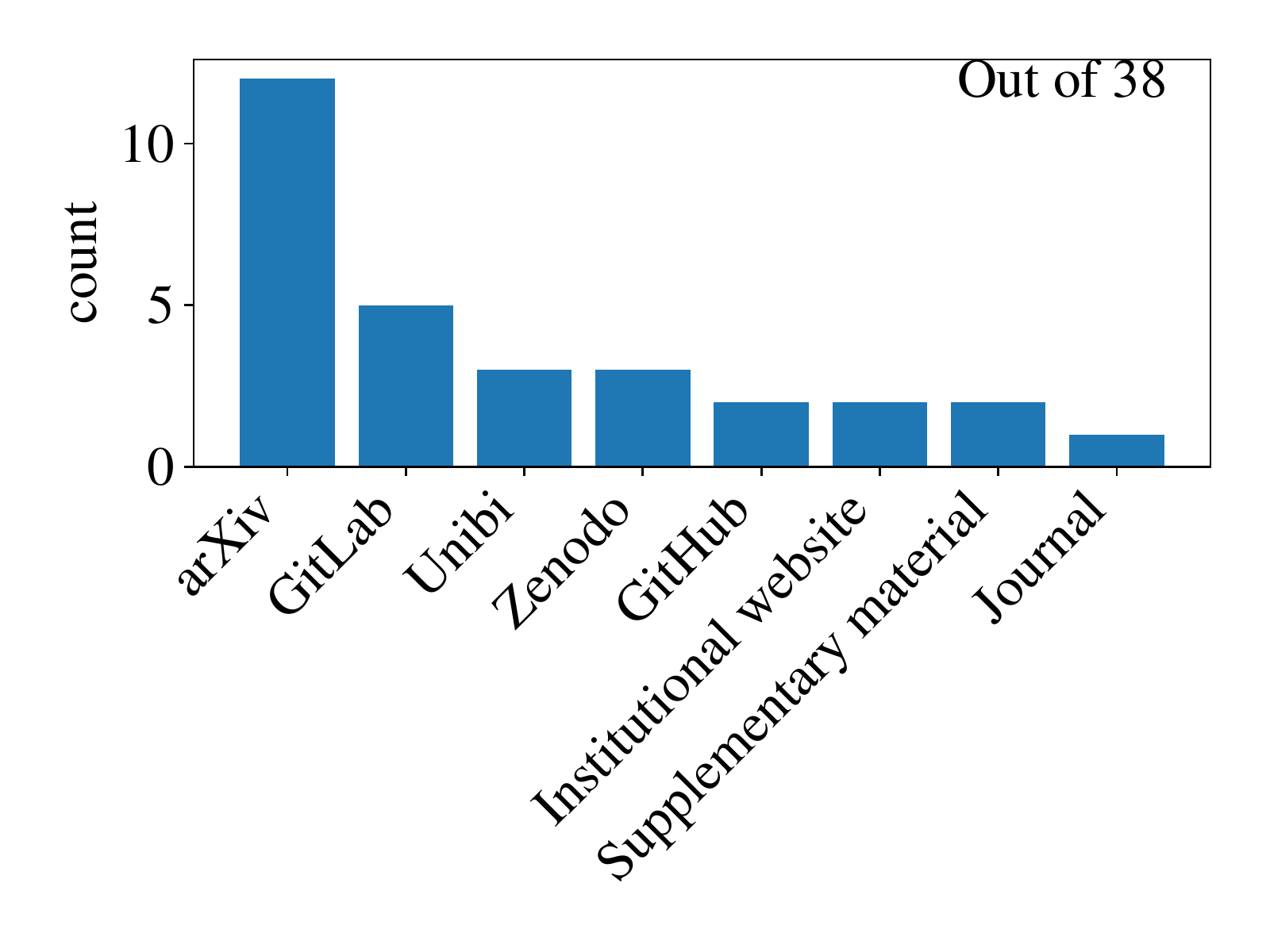}
  \captionof{figure}{\label{fig:data-repositories}Breakdown of the repositories
    used to host data published by the authors of the submissions indicated in
    Fig.~\ref{fig:publish-data}. Unibi is the institutional repository of the
    University of Bielefeld.}
\end{minipage}
\end{figure}

Where configurations are not generated as part of the work, they must be
obtained from somewhere. This may be a collaborations own storage, or
publicly-available ensembles may be used. In either case storage infrastructure
must be used, and in the latter case a facility for finding and sharing the
configurations. The International Lattice Data Grid (ILDG) provides schemas and
specifications for such services, which have been deployed in a number of
Regional Grids (RGs). This was a very early example of FAIR data sharing not
only in LFT but in all of academia; it in fact pre-dated the coining of the
``FAIR'' acronym itself. Being the first in the space has however meant that
tooling development in the surrounding world has moved on, and left the ILDG
tooling unmaintained. Only 14 of 230 submissions using existing configurations
acknowledge the hosting and sharing infrastructure used
(Fig.~\ref{fig:acknowledges-cfg-hosting}), all of which acknowledge the Japan
Lattice Data Grid (Fig.~\ref{fig:cfg-hosting-breakdown}), reflecting the dormant
state of the ILDG project and the RGs. Working groups have now however resumed
work on ILDG to modernise it, with the support of new funding, as discussed in
ILDG's joint contribution~\cite{ildg}.

Publications can also share data other than field configurations. Typically
being smaller in size, these require less infrastructure so can be shared with
generally-available data repositories such as CERN Zenodo. An advantage of using
such a repository over using a regular code or web hosting service is that they
will typically provide a persistent identifier (PID) such as a Digital Object
Identifier (DOI), which gives a reference that will not change over time
(e.g.\ when a researcher changes their GitHub username, or institution), and a
commitment to remain available in the moderately long term (unlike services such
as GitLab, which recently announced older repositories would be made unavailable
after a period of inactivity, with only a few weeks' notice\footnote{This
decision was later overturned due to public outcry, but there is no barrier to
the situation occurring again.}). Sharing data in this format, rather than
relying on tables in a PDF file, makes collating data from multiple sources and
re-using it in other work significantly easier, and removes the significant
likelihood of errors occuring when transcribing data.
Figure~\ref{fig:publish-data} illustrates that this is not yet common practice
in LFT, while Fig.~\ref{fig:data-repositories} shows a significant reliance on
code hosting services like GitHub and GitLab that do not provide persistent
identifiers or a guarantee of longevity. The phrase ``data available on
request'' is commonly used in some areas of science where sharing of data is
required by publisher policy but authors have not prepared it; in some cases
this has led to significant delays in access to data as requests are
ignored~\cite{supercooled-water}. This phrase has yet to catch on in LFT\@.

\begin{figure}
  \includegraphics[width=0.54\columnwidth]{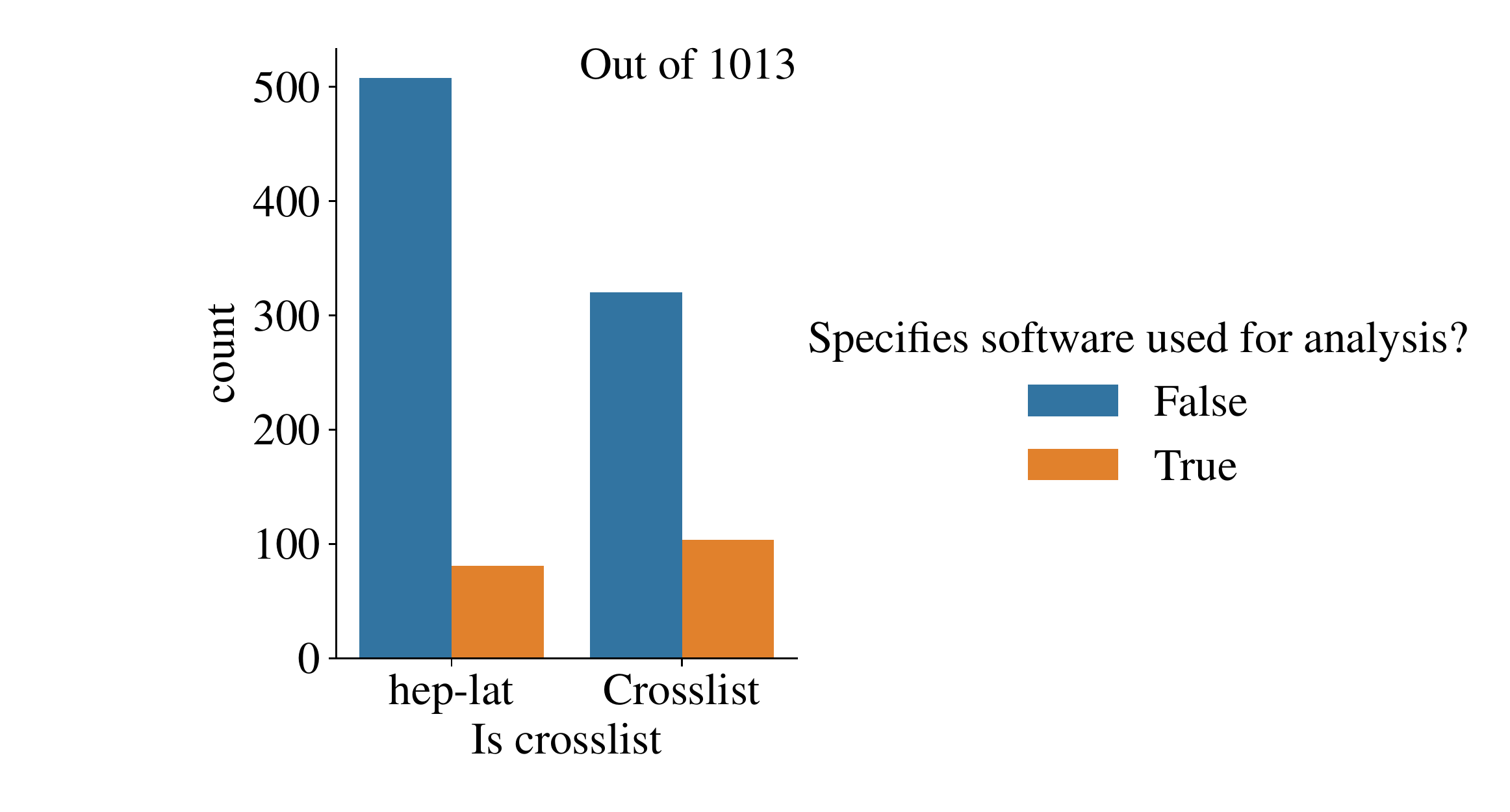}
  \hfill
  \includegraphics[width=0.44\columnwidth]{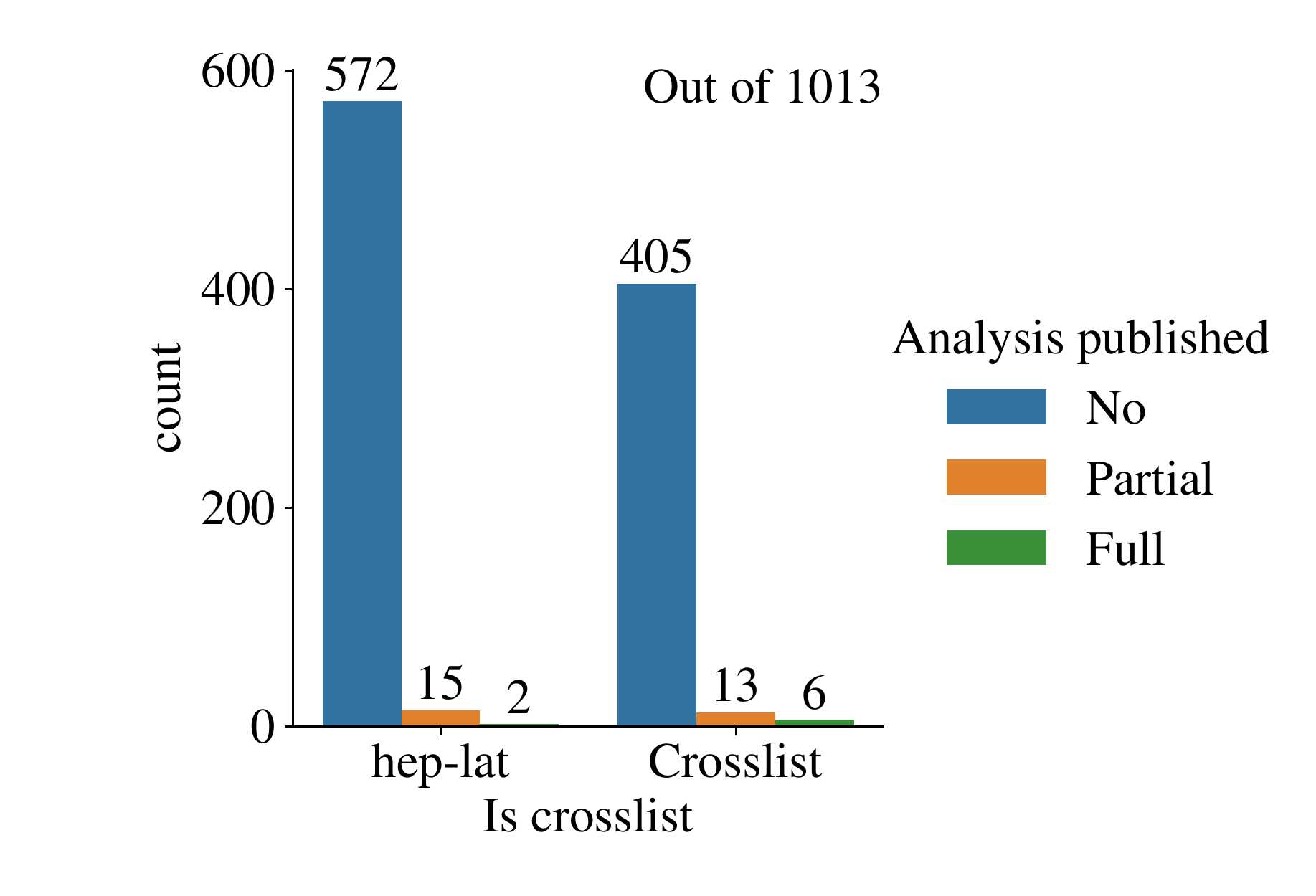}
  \caption{\label{fig:analysis-software}Breakdown of submissions to the hep-lat
    arXiv in 2021 that generate new numerical results that specify any of the
    software used for the analysis of the data presented (left panel), and those
    that publish the full or partial software workflow for this analysis (right
    panel).}
\end{figure}

\begin{figure}
  \centering
  \includegraphics[width=0.9\columnwidth]{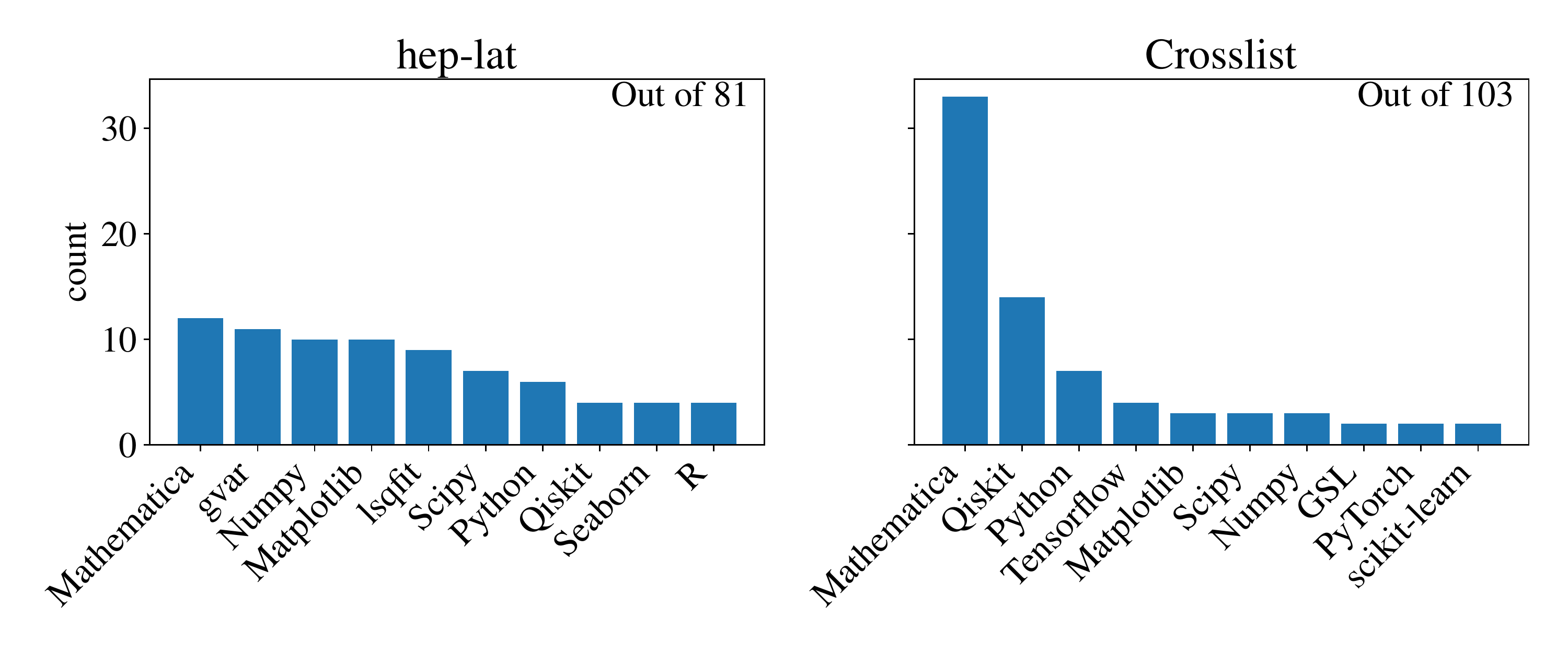}
  \caption{\label{fig:analysis-software-breakdown}Breakdown of the top ten
    specific pieces of  analysis software referred to by submissions to the
    hep-lat arXiv in 2021 that acknowledged such software.}
\end{figure}

The final part of the survey considers the data analysis of LFT data, i.e.~the
process that takes in the data produced from configurations on HPC facilities
and outputs the plots and tables shown in the submissions. Many areas of
experimental science have reproducibility efforts, and these have experimental
facilities in place of deterministically-generated field configurations, and
take measurements on these facilities that are not expected to give
machine-precision identical results when repeated; the reproducibility effort
focuses on ensuring that the data from experimental measurements always gives
the same numerical results once analysed. As discussed above, encoding these
procedures into software and sharing that software is the only feasible way to
do this unambiguously.

Figure~\ref{fig:analysis-software} illustrates that fewer than one in six
hep-lat numerical submissions specified any of the software used. The most
popular tools referred to are shown in
Fig.~\ref{fig:analysis-software-breakdown}; most are programming languages or
frameworks rather than computation-specific tools. Also shown in
Fig.~~\ref{fig:analysis-software} is that only two of these submissions included
the workflow allowing another researcher to fully reproduce the analysis
performed. This represents a significant opportunity for growth. A parallel
survey, the initial results of which are reported in a separate
contribution~\cite{andreas} indicates that significantly more work is enabled by
workflows that are at least partially automated; publication of these would
significantly boost the reproducibility of the data analysis phase of LFT
computations.

\section{Conclusions}

While LFT has in many cases been at the forefront of open science, being one of
the first disciplines to embrace open-access preprints and being ahead of its
time with FAIR data for field configurations, there are some areas that suffer
from a ``first-mover disadvantage'' where work is ongoing to realign with more
modern tooling, and others where there is an opportunity to learn from the
example of other disciplines.

Some low-hanging fruit to improve the reproducibility of publications include
specifying and acknowledging publicly-available software that has enabled the
research presented, and making existing software workflows available (and citing
them in work that uses them). Taking manual processes and automating them so
they can be published is more challenging, in particular doing so in a way that
can be run end-to-end without intervention. There is work to be done to write
tools that will enable this to be done more easily; this work will be informed
by a parallel survey of individual researchers' practices in this area, whose
initial results are reported in a separate contribution~\cite{andreas}.

\section{Acknowledgements}
This work has been funded by the UKRI Science and Technologies Facilities
Council (STFC) Research Software Engineering Fellowship EP/V052489/1. The author
acknowledges the support of the Supercomputing Wales project, which is
part-funded by the European Regional Development Fund (ERDF) via Welsh
Government. The author would like to thank the organisers of the UKLFT Annual
Meeting for the invitation to present on this topic, which inspired this work.
The author also thanks Julian Lenz for reviewing the draft of this contribution.

\bibliography{reprosurvey} \bibliographystyle{apsrev}
\end{document}